\documentclass[12pt]{iopart}

\usepackage{graphicx}

\newcommand{\jj}{\mathrm{i}}
\newcommand{\cc}{\mathrm{c}}

\newcommand{\VE}{\stackrel{\rightarrow}{E}}
\newcommand{\VH}{\stackrel{\rightarrow}{H}}

\newcommand{\VS}{\stackrel{\rightarrow}{S}}

\newcommand{\ee}{\mathrm{e}}
\newcommand{\dd}{\mathrm{d}}
\newcommand{\J}[2]{\mathrm{J}_{\mathrm{#1}}\left(#2\right)}

\newcommand{\K}[2]{\mathrm{K}_{\mathrm{#1}}\left(#2\right)}

\newcommand{\ddiv}{\mathrm{div}}

\newcommand{\RR}{\mathcal{R}}

\newcommand{\TA}{\tilde{A}}
\newcommand{\TB}{\tilde{B}}

\newcommand{\rr}{\tilde{\rho}}
	
\newcommand{\ccc}{\mathrm{c.c.}}

\newcommand{\eE}{\tilde{E}}
\newcommand{\eH}{\tilde{H}}

\newcommand{\DD}{D_{\mathrm{eff}}}
\newcommand{\chalco}{$\mathrm{As_2Se_3}$}

\begin{document}
\title[SBS revisited: Strong coupling regime and Rabi splitting]{Stimulated Brillouin scattering revisited: Strong coupling regime and Rabi splitting}


\author{Kien Phan Huy, Jean-Charles Beugnot, Jo\"el Cabrel Tchahame \& Thibaut Sylvestre}

\address{Institut FEMTO-ST, Universit\'e Bourgogne Franche-Comt\'e, Centre National de la Recherche Scientifique (CNRS), UMR 6174, Besan\c con, France}
\ead{kphanhuy@univ-fcomte.fr}
\vspace{10pt}
\begin{indented}
\item[]August 2015
\end{indented}

\begin{abstract}
Stimulated Brillouin scattering in optical waveguides is a fundamental interaction between light and acoustic waves mediated by electrostriction and photoelasticity. In this paper, we revisit the usual theory of this inelastic scattering process to get a joint system in which the acoustic wave is strongly coupled to the interference pattern between the optical waves. We show in particular that, when the optoacoustic coupling rate is comparable to the phonon damping rate, the system enters in the strong coupling regime, giving rise to avoided crossing of the dispersion curve and Rabi-like splitting. We further find that optoacoustic Rabi splitting could in principle be observed using backward stimulated Brillouin scattering in sub-wavelength diameter tapered optical fibers with moderate peak pump power.
\end{abstract}

%
%
%
%
%

\section{Introduction}
In the last few years, the field of cavity and Brillouin optomechanics, a branch of physics which focusses on the interaction between light and tiny mechanical and acoustical resonators, has drawn widespread interest because of key fundamental observations such as resolved-sideband cooling, optomechanically induced transparency, quantum coherent coupling, or Rabi-like oscillations \cite{Aspelmeyer,COM,Bahl2012,Bahl2015}. Rabi oscillations were initially described as damped periodic oscillations of an excited atom coupled to an electromagnetic cavity in which the atom alternately emits and reabsorbs photons \cite{Haroche}. From a theoretical point of view, this remarkable and ubiquitous phenomenon can be readily described as a joint system of two-coupled oscillators. If both oscillators are uncoupled they both share the same degenerate eigenfrequency. But when they are strongly coupled, the degeneracy is removed and the frequency is split in two distinguishable eigenfrequencies corresponding to the odd and even supermodes of the joint system. The frequency difference between the two eigenfrequencies is called the Rabi frequency \cite{Haroche,ExPolariton}. When the same effect occurs at the quantum level, meaning that the two oscillators are for instance an atom and a photon embedded in a cavity, the new eigensolutions are coherent superpositions of the two particles. When the two particles are not interacting with each other, they have their own dispersion relations $k_i(\omega)$ that may cross each other. However, if the strong coupling regime is achieved, the joint system dispersion curve displays an avoided crossing and Rabi splitting. Since its first observation, this concept has been extended to cavity optomechanics where mechanical and optical modes of high quality-factor resonators can be strongly coupled, and to many other joint systems including photons, phonons, excitons, polaritons or plasmons \cite{COM,Haroche,ExPolariton,SPP,PP}. 

In this work, we show that strong coupling regime and its formalism can also be applied to another strong photon-phonon interaction, which is known as stimulated Brillouin scattering (SBS).  Specifically, we theoretically demonstrate that such a strong coupling regime can be achieved between the acoustic phonon and the optical pump and Stokes waves using backward SBS in a subwavelength-diameter tapered optical fiber \cite{BeugnotOL2014,Beugnot2014}. Associated with the strong coupling is the observation of an avoided crossing and Rabi-like splitting between the optical and acoustical dispersion branches. 

The paper is organized as follows: First, we describe the principle and methodology of the stimulated photon-phonon interaction. Then we derive from the standard theory of SBS a joint two-level system, as commonly used in the Rabi problem. We investigate from this system the specific conditions that allow for the achievement of strong optoacoustic coupling regime in tapered optical fibers. Both silica and chalcogenide glass materials are compared with different taper diameters and optical pump powers. Finally, we demonstrate that the strong coupling regime and Rabi-like splitting could be observed in a sub-micron (0.7$\mu m$) chalcogenide fiber taper with moderate peak pump power (80 W).

\section{Principle and Methodology}
 \begin{figure}[htbp]
 \centering
 \includegraphics[width=0.6\columnwidth]{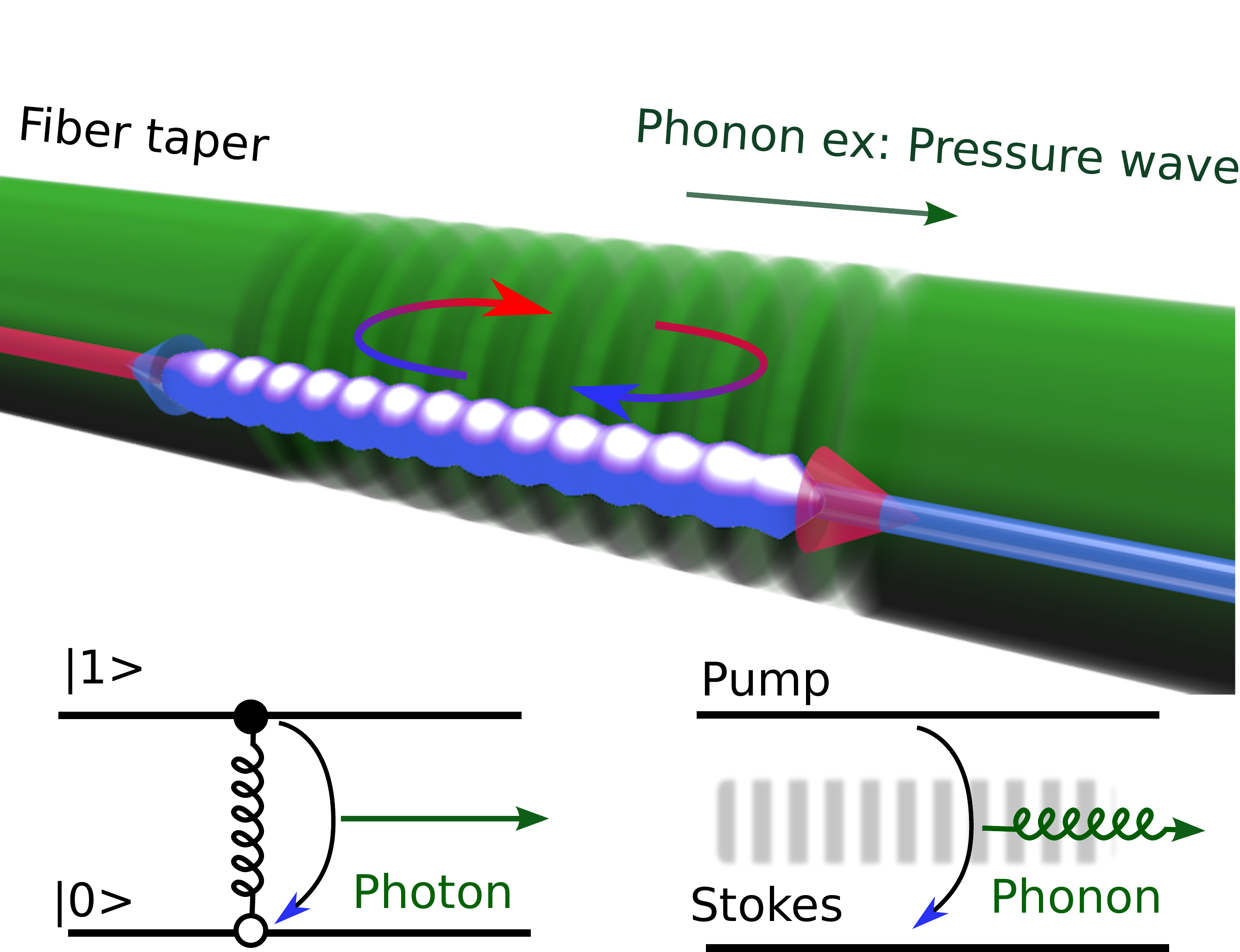}%
 \caption{Principle of strong coupling regime using backward stimulated Brillouin scattering in an optical fiber taper. A strong pump wave (red) coherently interacts with a counterpropagating probe wave, giving rise to an optical interference pattern which generates and couples to an acoustic wave. Under strong coupling regime, the interference pattern and the phonon hybridized and alternately exchange energy over period of few microns. \label{Fig1}}
 \end{figure}
Let us first describe the stimulated photon-phonon interaction under consideration. When coherent laser light is coupled and guided into an ultrathin optical fiber, as that shown schematically in Fig.~\ref{Fig1}, light generates and interacts with several types of acoustic waves \cite{Beugnot2014,Royer}. Here we assume only one acoustic pressure wave and we will find below the specific conditions to do so in a fiber taper. As scketched in Fig.~\ref{Fig1}, SBS is an inelastic scattering whereby two frequency-detuned optical pump (in red) and Stokes (in blue) waves coherently interact in a dielectric material, giving rise to an optical interference pattern (purple) which generates an acoustic wave (green) from the electrostrictive force. Simultaneously, the photoelastic effect creates an index grating that travels at the speed of hypersound (a few thousand of m.s$^{-1}$ in silica and chalcogenide glasses). Pump light then scatters in the backward direction on the index grating that acts as a Bragg mirror, with a Doppler downshift that corresponds to the frequency detuning between the two optical waves. When the frequency detuning is equal to the phonon frequency, phase-matching is achieved and thus one gets amplification of the Stokes signal. In the weak coupling regime, both the optical interference pattern contrast and acoustic wave grow along the optical waveguide, whereas in the strong coupling regime, they hybridized and alternately exchange energy over short period of a few microns, associated with the phonon lifetime.

From a quantum point of view, a pump photon is annihilated to create both a Stokes photon and an acoustic phonon, as depicted in bottom left of Fig.~\ref{Fig1}. Similarly to the Rabi problem, the two pump and Stokes photons can be seen as the two levels of an artificial atom. In our case however, the two coupled objects are actually the phonon and the optical beating between the pump and Stokes waves which are coupled by the electrostrictive force. As a result, our joint system is fundamentally different from the exciton polariton picture. In the exciton polariton case, the two level system is supposed to have only one particle. It is then described as an exciton with a given resonant frequency, as sketched by the \textit{spring} in bottom left of Fig.~\ref{Fig1}. The photon interacting with the exciton is in fact a signal probe tuned at the resonant frequency. The experiment usually consists in tuning the probe signal by adjusting its angle (wavevector) or changing the temperature to affect the resonant frequency of the exciton. In the SBS case, the pump and Stokes light replace the two levels since $\omega_P>\omega_S$ (see bottom right of Fig.~\ref{Fig1}). However, those two levels are not bounded by a resonant frequency as defined by the gap of a two-level atom. In SBS, the Brillouin frequency shift (BFS) is given by the phase-matching condition between the acoustic properties of the mechanical structure formed by the fiber taper and the optical waves. The resonant nature of the process mostly depends on the acoustic properties of the fiber. This is illustrated in bottom right of Fig.~\ref{Fig1}, where the phonon plays a role of \textit{spring}. Since the pump and Stokes waves are coherent laser lights that are conveniently produced and manipulated, we suggest to use them for probing the coupling with the acoustic phonon. 

\begin{figure}[hbt]
\centering
\includegraphics[width=0.8\columnwidth]{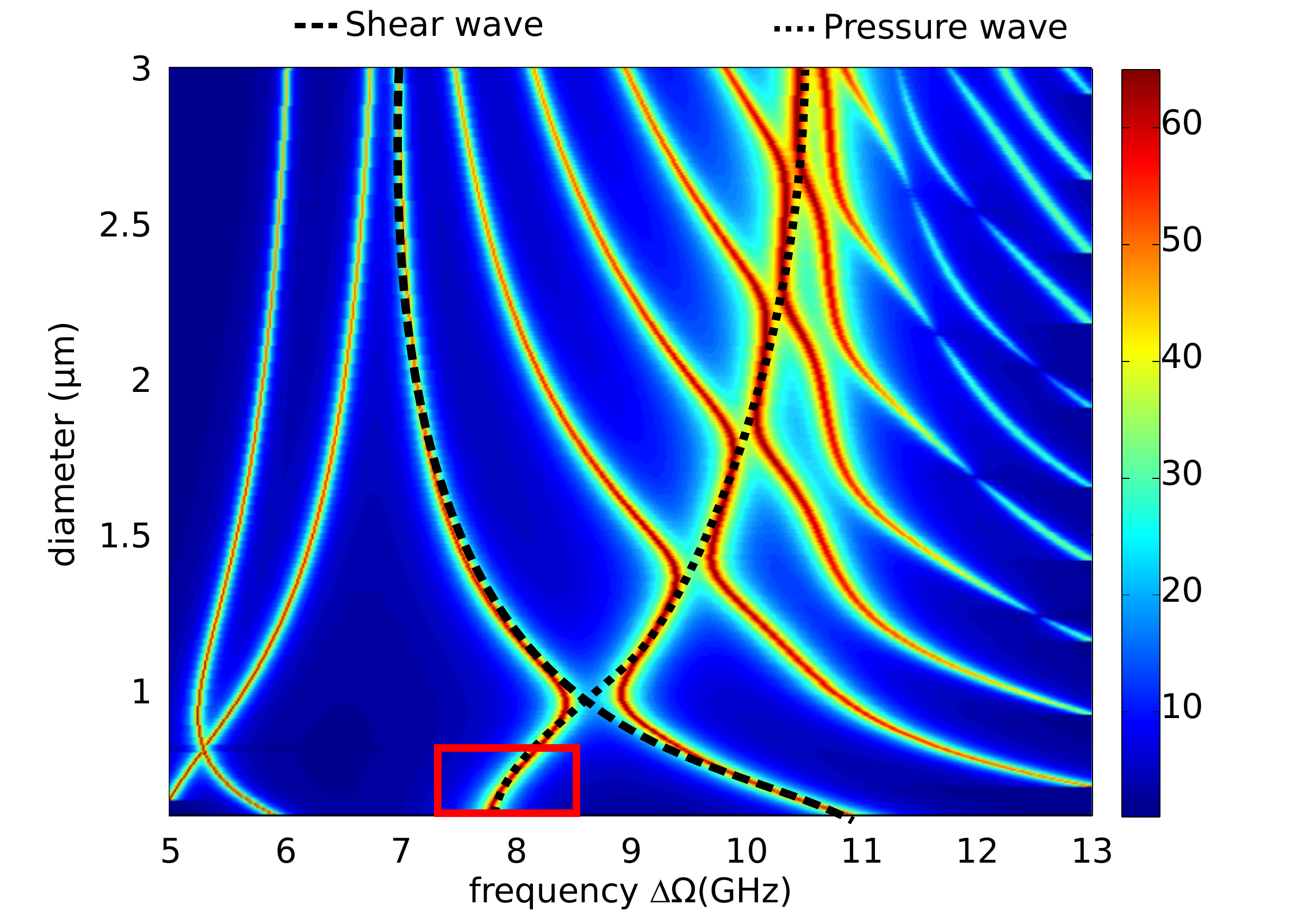}%
\caption{Color plot of the coupling efficiency between the acoustic modes with the fundamental optical mode of a silica fiber taper as a function of the acoustic frequency (horizontal axis) and of the taper diameter (vertical axis). Several avoided crossings appear when a pressure (P) wave (dotted) crosses a shear (S) wave (dashed). The squared red area isolates a single acoustic mode, required for the strong coupling regime. \label{Fig2}}
\end{figure}

To observe Rabi-like splitting in such a two-level system, it is also required to limit the optoacoustic interaction to ideally one guided acoustic mode, which is not possible using a bulky standard optical fiber with a 125-$\mu$m cladding. This can however be achieved using an optical fiber tapered down to a sub-micrometer diameter fiber, to get only a few discrete acoustic modes. To that end, we first solved the dispersion equation of acoustic modes $\beta_a(\Omega)$ in a silica rod as a function of the taper diameter. This dispersion relation reads as \cite{Royer} 
\begin{eqnarray}
	\frac{2p}{a}(q^2+\beta_a^2)\J{1}{pa}\J{1}{qa}-(q^2-\beta_a^2)^2\J{0}{pa}\J{1}{qa}=4\beta_a^2pq\J{1}{pa}\J{0}{qa}
\end{eqnarray}
with $p=\sqrt{\frac{\Omega^2}{V_L^2}-\beta_a^2}$ and $q=\sqrt{\frac{\Omega^2}{V_T^2}-\beta_a^2}$ where $V_L$ and $V_T$ are the longitudinal (P) and shear (S) acoustic velocities, respectively. $J_i$ are the Bessel functions. $a$ is the fiber taper radius. 
We then did the same work for the optical modes, as was done in Ref.\cite{Snyder}.  The optical dispersion equation for a step index fiber of  core refractive index $n_1$ and cladding refractive index $n_2$ is
    \begin{eqnarray}
       \left\{\frac{\mathrm{J}'_\mathrm{m}(\gamma_1a)}{\gamma_1a \J{m}{\gamma_1a}}
        +\frac{n_1^2}{n_2^2}\frac{\mathrm{K}'_{\mathrm{m}}(\gamma_2a)}{\gamma_2a \K{m}{\gamma_2a}}\right\}
        &=&\frac{m}{a^2}\left(\frac{1}{\gamma_1^2}+\frac{1}{\gamma_2^2}\right)
        \left(\frac{1}{\gamma_1^2}+\frac{n_2^2}{n_1^2}\frac{1}{\gamma_2^2}\right),
    \end{eqnarray}
    where $ \gamma_1=\sqrt{k_0^2n_1^2-\beta^2}$, $\gamma_2=\sqrt{\beta^2-k_0^2n_2^2}$, and $K_\mathrm{m}$ denotes the modified function of the second kind with the prime denoting differentiation with respect to the argument. $k_0$ is the propagating wave vector in vacuum. For each integer value $\mathrm{m}$, the eigenvalue $\beta$ is the effective propagation constant along the fiber axis of the given mode. This leads to the propagating constants of the pump (P) and Stokes (S) waves $\beta_P(\omega)$, and $\beta_S(\omega)$. Then the phase-matching condition $\beta_P(\omega+\Delta\Omega)-\beta_S(\omega)=\beta_a(\Delta\Omega)$ sets the detuning frequency $\Delta\Omega$ at which Brillouin scattering occurs. 
    Figure \ref{Fig2} typically shows a numerical simulation of the acoustic wave spectrum generated in a silica fiber taper, as a function of frequency detuning $\Delta \Omega$ and of the fiber diameter\cite{Royer,Beugnot2014}. The color plot that represents the overlap between acoustic and optical modes actually corresponds to the scattering efficiency. As can be seen, there are several types of acoustic waves including longitudinal, shear, and surface waves around 5 GHz \cite{Beugnot2014}. We can also see several avoided crossings in Fig.~\ref{Fig2}, that appear when a pressure (P) wave (dotted) crosses a shear (S) wave (dashed). Their polarizations are actually no longer orthogonal and thus they can interact with each other, giving rise to such avoided crossing. We must stress here that they are not related with the strong coupling regime described in the next section.  Note also that around 5 GHz there is no anticrossing between the two surface Rayleigh waves because they have orthogonal polarizations. The squared red area in Fig.~\ref{Fig2} shows that, for sub-wavelength diameter fiber, there is a small frequency range where we can isolate a single acoustic mode around 8 GHz. We will investigate the strong coupling regime in this area, which is sufficiently far from avoided crossing.
    
Additionally, to reach the strong coupling regime, it is also needed to have a sufficiently strong coupling strength to transfer the energy back and forth between the optical beating and the phonon. As a result, large optical pump power is required. We will see thereafter that chalcogenide glass-based fiber tapers with large Brillouin gain are the best potential candidates enabling the strong coupling regime. Also, it is important to stress here that we will not consider the radiation pressure in our model because it is negligible for the optical fiber tapers under investigation. Recent works have however shown that Brillouin gain could be significantly enhanced in nanoscale and large step index waveguides when radiation pressure enters into play or becomes comparable to the electrostrictive force \cite{Rakich1,Rakich2}. In our model, we will also neglect the anti-Stokes scattering and the second-order Stokes scattering \cite{Boyd}. That assumption will be discussed later in \ref{A5}.
   
\section{Theory}
In the following, we reformulate the usual theory of SBS, initially derived by Boyd \cite{Boyd}, to yield a joint system that can experience Rabi splitting. The pump ($P$) and Stokes ($S$) fields and the material density $\rr$ can be expressed as,
\begin{eqnarray}
	\tilde{E_{P}}(x,y,z,t)&=&\TA_{P}(z,t)E_{P}(x,y)\ee^{\jj(\beta_{P} z-\omega_{P} t)}+\ccc, \label{eq:P}\\
	\tilde{E}_{S}(x,y,z,t)&=&\TA_{S}(z,t)E_{S}(x,y)\ee^{\jj(\beta_{S} z-\omega_{S} t)}+\ccc, \label{eq:S}\\
	\tilde{\rho}(x,y,z,t)&=&\rho_0+\left\{\rr(z,t)\RR(x,y)\ee^{\jj(\beta_a z -\Omega_a t)}+\ccc \right\}. \label{eq:A}
\end{eqnarray}
where $\rho_0$ is the mean density of the medium, $\beta_{P,S,a}$ and $\omega_{P,S}$ and $\Omega_a$ are the wavevectors and frequencies and acoustic wave, respectively. $\TA_{P,S}$ , $\rr(z,t)$ are slow varying enveloppes and $E_{P,S}$ and $\RR$ are the transverse profiles of each wave. Note that the P-wave model for an acoustic wave in a submicron fiber taper is appropriate if the frequency detuning range surrounds a P-wave branch and is far enough from the anticrossing with S-wave branch, as shown in the squared area in Fig.~\ref{Fig2}. In this area, we can readily assume that only one acoustic wave can be possibly coupled to the two optical waves. Introducing these equations into the wave equation and omitting the complex conjugate, we find \cite{Boyd}
  \begin{eqnarray}  
	\frac{\partial \TA_{P} }{\partial z}
	 +\frac{\partial \beta}{\partial \omega}\frac{\partial \TA_{P}}{\partial t}
	 &=& \frac{\jj \omega_P\gamma_e}{2n_P \cc \rho_0} \frac{1}{\DD}\rr\TA_S,\label{eq:Y1}\\
	-\frac{\partial \TA_{S} }{\partial z}
	 +\frac{\partial \beta}{\partial \omega}\frac{\partial \TA_{S}}{\partial t}
	 &=&\frac{\jj \omega_S\gamma_e}{2 n_S \cc\rho_0}\frac{1}{\DD}\rr^*\TA_P,\label{eq:Y2}
  \end{eqnarray} 
where $\ee^{\jj(\beta_k z-\omega_k t)}$, $\gamma_e$ is the electrostrictive constant, $n_P$, $n_S$ are the refractive indexes for the Pump and Stokes waves respectively, $\cc$ is the speed of light in vacuum, $\DD$ is a factor coming from normalization that is discussed in \ref{A1}. Note that this set of equations describes the nonlinear propagation of P and S waves but it does not exhibit the optical interference pattern. To that end, we need to combine them in order to get the equation of the optical beating.

\subsection{Poynting vector: Optical beating}
Since the optical beating is a local intensity modulation induced by the superposition of P and S waves, it can be described by the z component of the Poynting vector
  \begin{eqnarray}  
	\VS&=&\VE \times \VH,\\
	&=& \VS_P+\VS_S+\VE_P \times \VH_S+\VE_S \times \VH_P,\label{eq:Poynting}
  \end{eqnarray} 
where $\VS_{P,S}$ are the Poynting vectors for the Pump and Stokes wave. The two first terms of the right-hand side (RHS) describe the independent $S$ and $P$-wave Poynting vectors. They are linked to power evolution by the flux $\Phi$. The two last terms in the RHS displays the coherent local intensity modulations due to the interference. Those terms do not describe a power flow since they vanish after integration, however they describe a local intensity fluctuation, i.e., the optical beat note. Using the expressions of the Pump and Stokes waves, (\ref{eq:P}) and (\ref{eq:S}), we can readily see that this intensity fluctuations varies as $\ee^{\jj(\Delta \beta -\Delta\omega t)}$, where $\Delta \beta=\beta_P-\beta_S$ and  $\Delta \omega=\omega_P-\omega_S$. The optical beating thus travels at a speed defined by $v_b=\frac{\Delta \omega}{\Delta \beta}$. It is around a few thousands m/s, the same speed of the acoustic wave, if phase-matching is satisfied. In the following we will transform equations (\ref{eq:Y1}) and (\ref{eq:Y2}) to derive an equation of the optical beating and its relationship with the phonon. Careful attention was paid to the omitted complex conjugate terms, however all the terms of equation (\ref{eq:Poynting}) are not detailed and the focus is put only on phased-matched terms.
First, we introduce the amplitude of the magnetic field $\TB_P$. In the plane wave approximation, that gives for the P-wave $\TB_P(z,t)=-\frac{\beta_P}{\omega_P\mu_0}\TA_P(z,t)$. Equations  (\ref{eq:Y1}) and (\ref{eq:Y2}) yield
  \begin{eqnarray}  
	\frac{\partial \TB_{P} }{\partial z}
	 +\frac{\partial \beta}{\partial \omega}\frac{\partial\TB_{P}}{\partial t}
	 &=& \frac{\beta_P}{\beta_S}\frac{\jj\omega_P\gamma_e}{2n_P \cc \rho_0}\frac{1}{\DD}\rr\TB_S, \label{eq:Y3}\\
	 \frac{\partial \TB_{S}}{\partial z}
	  -\frac{\partial \beta}{\partial \omega}\frac{\partial\TB_{S}}{\partial t}
	 &=& -\frac{\beta_S}{\beta_P}\frac{\jj\omega_S\gamma_e}{2n_S\cc\rho_0}\frac{1}{\DD}\rr^*\TB_P.\label{eq:Y4}
  \end{eqnarray}  
Multiplying equation (\ref{eq:Y1}) by $\TB_S^*$ and the conjugate of equation (\ref{eq:Y4}) by $\TA_P$, we find
  \begin{eqnarray*}  
	\frac{\partial \TA_{P} }{\partial z}\TB_{S}^*
	 +\frac{\partial \beta}{\partial \omega}\frac{\partial \TA_{P}}{\partial t}\TB_{S}^*
	 &=& \frac{\jj \omega_P\gamma_e}{2n_P \cc \rho_0} \frac{1}{\DD}\rr\TA_S\TB_{S}^*,\\
	  \TA_{P}\frac{\partial \TB^*_{S}}{\partial z}
	  -\frac{\partial \beta}{\partial \omega} \TA_{P}\frac{\partial\TB_{S}^*}{\partial t}
	 &=& \frac{\beta_S}{\beta_P}\frac{\jj\omega_S\gamma_e}{2n_S\cc\rho_0}\frac{1}{\DD}\rr \TA_{P}\TB_P^*.
  \end{eqnarray*} 
 Since $P$ and $S$ lights are propagating in forward and backward directions and they are only shifted in frequency by a few GHz, we can make the assumptions $ \frac{\beta_S}{\beta_P}\simeq -1$ and $n=n_S\simeq n_P$. Taking the sum of above equations, and neglecting the group velocity as explained in \ref{A2} we readily find
    \begin{eqnarray}  
	\frac{\partial \TA_{P}\TB_{S}^* }{\partial z}
	 &=&
	  -\frac{\jj \omega_P\omega_S\gamma_e\rr}{n \cc \rho_0\DD} 
	 \left(\frac{\TA_P\TB_{P}^*}{\omega_P}-\frac{\TA_S\TB_{S}^*}{\omega_S}\right).
	 \label{eqT}
  \end{eqnarray} 
On the RHS of equation \ref{eqT}, we see that the optical beat note is clearly coupled with the phonon $\rr$. The term within brackets is usually referred as the ``population inversion'' in a two-level system problem. Indeed, if both waves were propagating in the same direction, each $\frac{\TA_k\TB_k^*}{\omega_k}$ term would be proportional to the power $P_k$ of the related k-wave, and be denoted the ``population'' of level $k$. However in the specific case of \textit{counter propagating waves}, the name ``population inversion'' is not appropriate. Since each ``population'' term is proportional to its Poynting vector, its sign depends on the direction of the power flow along the $z$ axis. If the S-wave propagates in the backward direction (SBS), the flux of the Poynting vector for the Stokes wave is negative leading to $\frac{2\TA_S\TB_{S}^*}{\omega_S}=-\frac{P_{S}}{\omega_S}$. As a result, the term in brackets of the RHS of the equation should be named ``total population''. This situation is very specific to the study presented here, and that it does not occur in co-propagating or static cases like the well known atom-photon coupling. In the atom-photon picture, the conservation law stipulates that the sum of the populations is constant, $\dot{\rho}_{11}+\dot{\rho}_{22}=0$\cite{Boyd}. However, for counterpropagating waves P and S, the Poynting picture shows that the conservation turns to $\dot{\Phi}_P-\dot{\Phi}_S=0$ that is unusual.

In order to build a joint-system where the optical beating is coupled with the phonon, it is important that the coupling rate does not vary, in other words, to conserve a constant ``total population''. To fulfill this criterion, we have to assume that anti-Stokes and second-order Stokes scattering remain negligible (see \ref{A5}). Under this assumption, we can multiply equation (\ref{eqT}) with the previously omitted fast oscillating phase term $\ee^{\jj(\Delta \beta -\Delta\omega t)}$, add and subtract $-\jj\Delta\beta\eE_P\eH_S^*(z,t)$, simplify with equation \ref{eq:P} and take the Fourier transform, we get
          \begin{eqnarray}  
	\frac{\partial  \widehat{\eE_P\eH_S^*}(z,\Omega)}{\partial z}
	-\jj\Delta\beta
	\widehat{\eE_P\eH_S^*}(z,\Omega)
	&=&- \frac{\jj\omega_P\omega_S\gamma_e}{4 n\cc \rho_0\DD}
	 \left\{
	\frac{ P_P}{\omega_P}
	+ \frac{P_S}{\omega_S}
	\right\}	 
	\hat{\rho}(z,\Omega), \label{eq:obeat}
     \end{eqnarray}
          with $ \eE_P\eH_S^*(z,t)=\TA_{P}\TB_{S}^*(z,t)\ee^{\jj (\Delta \beta z-\Delta\omega t)}$, and the $hat$ denoting the Fourier transform. This equation actually describes the optical beat note in the Fourier domain, with $\Omega$ being the frequency detuning between Pump and Stokes waves. If the RHS is neglected, meaning that there is no nonlinear interaction, this equation becomes the dispersion equation of the optical beating. We can then see that the optical beating travels with the wavevector $\Delta\beta$ that is related to the speed $v=\frac{\Delta \omega}{\Delta\beta}$. 
    
\subsection{Acoustic equation}
To build the other half of the joint system, a similar equation must be derived for the acoustic phonon. Using a standard description of a pressure wave for the acoustic model, we get
        
         \begin{eqnarray} 
	-2\jj\Omega\frac{\partial \rr}{\partial t}
	+\left(\Omega_a^2-\Omega^2-\jj\Omega\Gamma_B\right)
	\rr
	-2\jj\beta_a v_a^2
	\frac{\partial \rr}{\partial z}
	&=&
	-\frac{\gamma_e\beta_a^2}{n\cc}\frac{1}{\DD}\TA_{P}\TB^*_{S}.  \label{eq:ph}	
     \end{eqnarray}     
The first and third terms of the left-hand side (LHS) of the equation show that the phonon travels at the speed $v_a=\frac{\Omega_a}{\beta_a}$. The second term of the LHS is the signature of the mechanical stresses induced by the phonon. It is zero at the resonant frequency $\Omega_a$. The RHS of Eq. (\ref{eq:ph}) provides the coupling rate with the optical beating. To write this equation, we have to suppose that the phase-matching for backward SBS is satisfied and therefore that the acoustic wavevector is twice the optical one, as $\beta_a=2\beta_P$. Now, we should describe the whole dispersion curve and not restrict our study to the phase-matching case only. We will thus not make this assumption and let $\beta_a(\Omega)$ varying in the surrounding of  frequency resonance ($\Omega-\Omega_a\ll \Omega_a$). The $\Omega_a^2-\Omega^2-\jj\Omega\Gamma_B$ term then reduces to $-\jj\Omega\Gamma_B$, which stands for the acoustic phonon losses.

In order to see the dispersion equation of the phonon, we will rewrite the above equation by keeping the fast oscillating phase term $\ee^{\jj( \beta_a z -\Omega_a t)}$. We then add and subtract $\jj\beta_a \rr\ee^{\jj(\beta_a z -\Omega_a t)}$ and straightforwardly get
\begin{eqnarray}
	\left\{
	-\jj\beta_a\rr+
	\frac{\Omega}{\Omega_a}\frac{1}{ v_a}\frac{\partial \rr}{\partial t}
	\right\}
	\ee^{\jj( \beta_a z -\Omega_a t)}
	+\frac{\partial \rr_\phi}{\partial z}
	&\simeq&
	\jj u(\Omega)
	\rr_\phi
	+\jj k_{po}\eE_{P}\eH^*_{S}, \label{eq:PHON_TD}
\end{eqnarray}	
where $\rr_\phi=\rr(z,t)\ee^{\jj( \beta_a z -\Omega_a t)}+\ccc$ and
\begin{eqnarray}	
	u(\Omega)&=&\frac{\jj\Omega\Gamma_B}{2\beta_a v_a^2},\label{eq:u}\\
	k_{po}&=&-\frac{1}{2\beta_a v_a^2}\frac{\gamma_e\beta_a^2}{n_S \cc}\frac{1}{\DD},
\label{eq:kpo}
\end{eqnarray}	
  represent the acoustic losses and the nonlinear coupling rate, respectively. The dispersion equation can be derived from the Fourier transform of equation \ref{eq:PHON_TD}. Making the assumption that the detuning from the Brillouin frequency shift is low $\Omega-\Omega_a\ll\Omega_a$, we find
  
\begin{eqnarray}
	\frac{\partial \hat{\rho}(z,\Omega)}{\partial z}
	-\jj\beta_a(\Omega)
	\hat{\rho}(z,\Omega)
	&=&
	\jj u(\Omega)
	\hat{\rho}(z,\Omega)
	+\jj k_{po}\widehat{\eE_P\eH_S^*}(z,\Omega),\label{eq:PHONON}
\end{eqnarray}	  
  where $\beta_a(\Omega)=\beta_a +(\Omega -\Delta\omega)\frac{1}{ v_a}$ is the dispersion relation of the phonon and 
$\rho(z,\Omega)$ is the Fourier transform of $\rr_\phi$.

\subsection{The joint system}
\begin{figure}[hbt]
  \centering
 \includegraphics[width=0.8 \columnwidth]{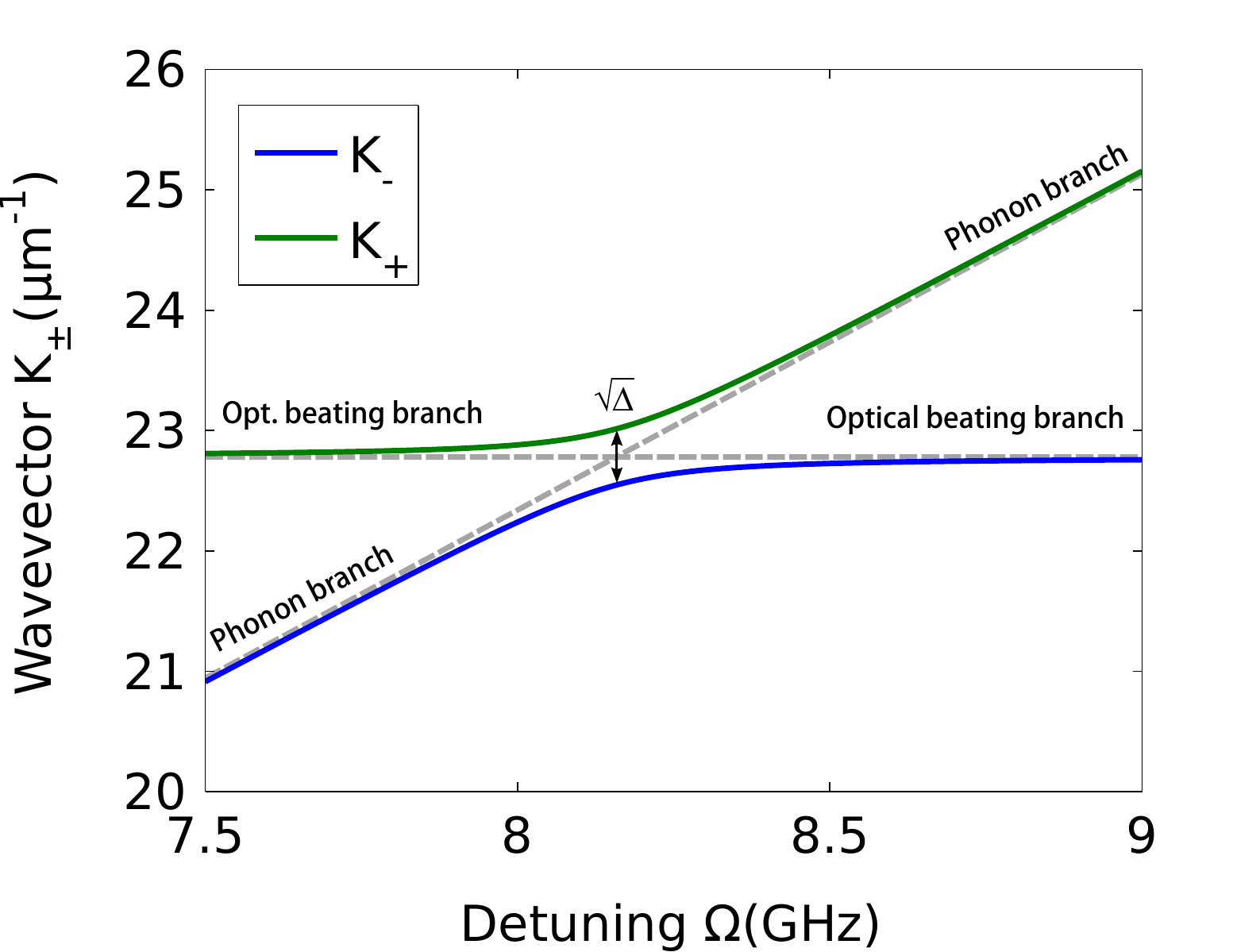}%
 \caption{Dispersion curves $K_{\pm}(\Omega)$ showing acoustic Rabi splitting and avoided crossing in a 0.7~$\mu$m mode-field diameter \chalco ~tapered optical fiber with 100~W pump power and 1~mW Stokes power. The parameters are from table \ref{tab1} of \ref{A4} and the phonon losses were neglected \cite{Chalco}. Dashed gray lines show the intersecting dispersion curves when there is no coupling. \label{Fig3}}
 \end{figure}

We can write equations \ref{eq:obeat} and \ref{eq:PHONON} as a joint system with $2\widehat{\eE_P\eH_S^*}$ denoting the optical beating 
                    
	\begin{eqnarray}  
	           \frac{\partial }{\partial z}
	           \left(
                    \begin{array}{c}
                    2\widehat{\eE_P\eH_S^*}\\
                   \hat{\rho}
                    \end{array}
                    \right)
                    &=&
                    \jj \left(
                    \begin{array}{cc}
		k_{oo} & k_{op}\\
		k_{po} & k_{pp}\\		
                    \end{array}\right)
                    \left(
                    \begin{array}{c}
                    2\widehat{\eE_P\eH_S^*}\\
                    \hat{\rho}
                    \end{array}
                    \right),\label{eq:joint}
     \end{eqnarray}  
where $k_{oo}=\Delta\beta$ and $k_{pp}=\beta_a(\Omega)+u(\Omega)$ are the eigenvalues of the uncoupled system. Those are the wavevectors of the optical beating and of the phonon, respectively. Note that in the expression of $k_{pp}$, we have
$u(\Omega_a)=\jj\frac{\Gamma_B}{2 v_a}$ that is related to the Brillouin linewidth $\Gamma_B$, and thus the phonon decay rate or lifetime.  Since this term is imaginary, it gives an imaginary part to $k_{pp}$ that stands for the phonon propagation losses. 
The coupling coefficient are  $k_{po}$, previously defined in equation (\ref{eq:kpo}), and $k_{op}=- \frac{\omega_P\omega_S\gamma_e}{2 n\cc \rho_0\DD} \left\{\frac{ P_P}{\omega_P}+ \frac{P_S}{\omega_S}\right\}$. The eigenvalues $K_{\pm}$ of the joint system are solutions of characteristic polynomial
\begin{eqnarray}  
	\left(K_{\pm}-k_{oo}\right)\left(K_{\pm}-k_{pp}\right)-k_{op}k_{po}&=&0.
\end{eqnarray}  
If $k_{op}k_{po}=0$, the coupling rate is null and there are two independent eigenvalues with dispersion curves that cross each other. If the coupling term is non zero, it is more convenient to write the equation as the form
\begin{eqnarray}  
	K_{\pm}^2-(k_{oo}+k_{pp})K_{\pm}+k_{pp}k_{oo}-k_{op}k_{po}&=&0,
\end{eqnarray}  
with
\begin{eqnarray}  
%
k_{op}k_{po}&=& \frac{\beta_a\Gamma_B\cc}{8n v_a}
	\frac{g_0}{A_{eff}}
	 \left\{\frac{ P_P}{\omega_P}+ \frac{P_S}{\omega_S}\right\}. 
\end{eqnarray}  
where $g_0$ 
is the Brillouin gain defined in \ref{A3} and $A_{eff}=\DD^2$ the effective area of the mode-field diameter (MFD). This equation possesses two solutions if the discriminant is positive
\begin{eqnarray}  
	\Delta=(k_{oo}+k_{pp})^2-4\left(k_{pp}k_{oo}-k_{op}k_{po}\right)&\ge&0.
\end{eqnarray}  
At the crossing of two dispersion curves ($\Omega=\Omega_a$), where the phase matching condition is fulfilled $\Delta\beta=\beta_P-\beta_S=\beta_a$, neglecting the phonon losses leads to $k_{oo}= k_{pp}$. We can rewrite this equation as
\begin{eqnarray}  
	\Delta=(k_{oo}-k_{pp})^2+4k_{op}k_{po}&\ge&0. \label{eq:condition}
\end{eqnarray}  
Here we see that, at the crossing point of the uncoupled dispersion curves, where the real part of $k_{oo}$ equals the real part of $k_{pp}$, only the coupling term and the imaginary parts of $k_{oo}$ and $k_{pp}$ still remains in the equation. 
%
%
Therefore, if the sum of the remaining terms is positive the eigenvalues are
\begin{eqnarray}  
	K_{\pm}&=&\frac{1}{2}\left\{
	k_{oo}+k_{pp}\pm \sqrt{\Delta}
	\right\}.\label{eq:Eigenwavevector}
\end{eqnarray}  
We first compute the result of this equation, neglecting the phonon losses $u(\Omega)=0$. This is illustrated in Fig.~\ref{Fig3} that shows the wavevector $K_{\pm}$ as a function of detuning $\Omega$ for a 0.7~$\mu$m mode-field diameter chalcogenide tapered optical fiber, with 100~W and 1~mW pump and Stokes power. 
The Rabi splitting is clearly visible in Fig.~\ref{Fig3} as an avoided crossing between the optical beating and phonon branches. Specifically, the splitting is characterized by a strong gap between the two curves of $\sqrt{\Delta}$. As a comparison, the dashed gray lines show the intersecting dispersion curves when there is no coupling. Note that for the simulation shown here, we considered an acoustic velocity of 2250~m.s$^{-1}$ for chalogenide glass \cite{BeugnotOL2014}, so the splitting occurs around 8~GHz. Now, we will take into account the phonon losses and see how it will affect the Rabi splitting. Furthermore, we will define a criterion for the strong coupling regime and study how this criterion evolves with the material nonlinearity, the pump power and the fiber taper diameter.
 
\section{Phonon losses and splitting ratio}

In the last paragraph, the imaginary part of $k_{oo}$ and $k_{op}$ were neglected making each wave lossless. In the exciton-polariton picture, the weak and strong coupling regime are usually separated by comparing the coupling rate $g$ and the exciton-photon decay rates $\gamma_{ph}$, and $\gamma_{ex}$. If  $g\gg \gamma_{ex},\gamma_{ph}$ the system is in the strong coupling-regime whereas if $g \ll \gamma_{ex},\gamma_{ph}$ the system is in the weak coupling regime \cite{CQED}. Our case using SBS is very similar. If we do not neglected the phonon losses in (\ref{eq:condition}), we get the following inequality 
\begin{eqnarray}  
	\sqrt{\Delta}=2\sqrt{k_{op}k_{po}}&\ge&|u(\Omega)|.\label{eq:crit}
\end{eqnarray}  
where the right-hand side $\sqrt{\Delta}$ plays a role similar to the coupling rate $g$, with the difference that our eigenvalues are wavevectors, and $\sqrt{\Delta}$ is thus homogeneous to $m^{-1}$, instead of $s^{-1}$. The LHS term $u(\Omega)$ defined by equation (\ref{eq:u}) is related to the phonon decay rate, $\gamma_{ex}$ in the exciton-polariton picture. To discriminate the weak coupling from the strong coupling regime, it is convenient to rewrite equation (\ref{eq:crit}) with the splitting ratio $p=\frac{\sqrt{\Delta}}{|u(\Omega)|}$. If $p\ll1$, the joint system is in the weak coupling regime whereas if $p\gg 1$, it is in the strong optoacoustic coupling regime.

\begin{figure}[ht]
  \centering
 \includegraphics[width=1 \columnwidth]{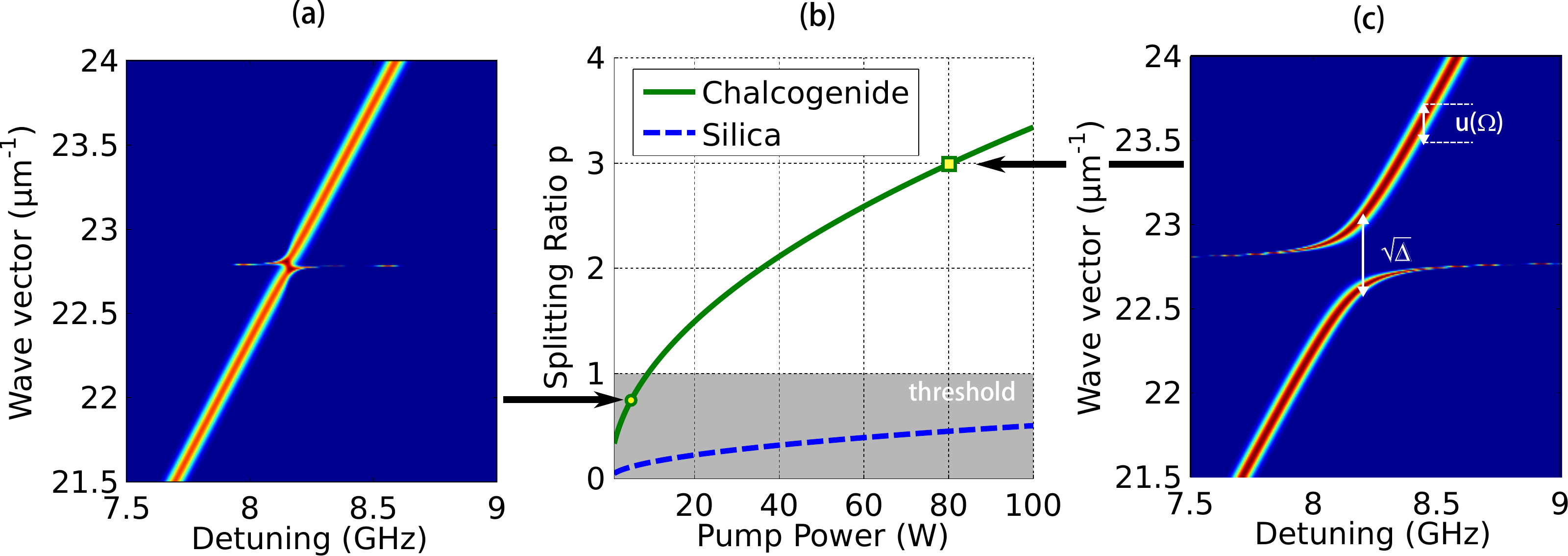}%
 \caption{(a) Dispersion curves $K_{\pm}(\Omega)$ for 5~W Pump power and 1~mW Stokes power in a 0.7~$\mu$m mode-field diameter tapered \chalco ~fiber. (b) Wavevector splitting ratio $p$ \chalco~ and silica fiber taper versus the pump power. $p$ measures the gap between two split curves in number of linewidth. The gray area delimits the weak coupling regime. (c) Dispersion curve $K_{\pm}(\Omega)$ for 80~W Pump power and 1~mW Stokes power. Optical loss is set to 1 dB.m$^{-1}$.\label{Fig4}}
 \end{figure}

Figure~\ref{Fig4} compares these two distinct regimes for an 0.7 $\mu$m-diameter optical fiber taper based on either chacolgenide or silica glass materials, whose parameters are listed in \ref{tab1} in \ref{A4}. The chosen glass materials are silica and chalcogenide \chalco ~glasses because they are available as micro and nanowires \cite{Beugnot2014,Baker2012}. This comparison allows us to show the significant role of the Brillouin gain in the splitting ratio. Figure~\ref{Fig4}(a) shows in particular the case where $p<1$, for a pump power of 5W, in the gray area of Fig. \ref{Fig4}(b). In this case, the nonlinear process is not strong enough and no splitting can be observed. Figure~\ref{Fig4}(c) clearly shows instead the strong gap $\sqrt{\Delta}$ between the two branches for a pump power of 80 W. Rabi splitting occurs when the gap then exceeds a certain number of Brillouin linewidths defined by $|u(\Omega)|$.

This number of linewidths is indeed the splitting ratio $p$ that is plotted in Fig.~\ref{Fig4}(b) as a function of the pump power. One can see that, for a the silica-based taper, the Brillouin gain is not strong enough to induce any splitting even at high pump power, whereas the chalcogenide fiber taper enables the splitting from a pump power just above 10~W. This is simply due to the the fact that Brillouin gain in \chalco~ fiber taper is almost 200 times that of silica (See Table \ref{tab1}). The splitting ratio is further illustrated in Fig.~\ref{Fig4}(c) for 80 W power. In this case $p=3$, we thus expect a gap between the two branches of 3 linewidths.

A further comparison between Figs.~\ref{Fig3} and \ref{Fig4} shows that the optical beating branch vanishes. This is due to the fact that optical beating experiences negligible loss (1 dB.m$^{-1}$), and thus its branch is very thin and beyond image resolution. Unlike the optical beating, acoustic phonon experiences much larger loss, leading to a much wider branch than the optical one. Consequently, we can deduce that the resonant nature of the process is definitely related to the phonon lifetime.
    \begin{figure}[hb]
  \centering
 \includegraphics[width= 0.8\columnwidth]{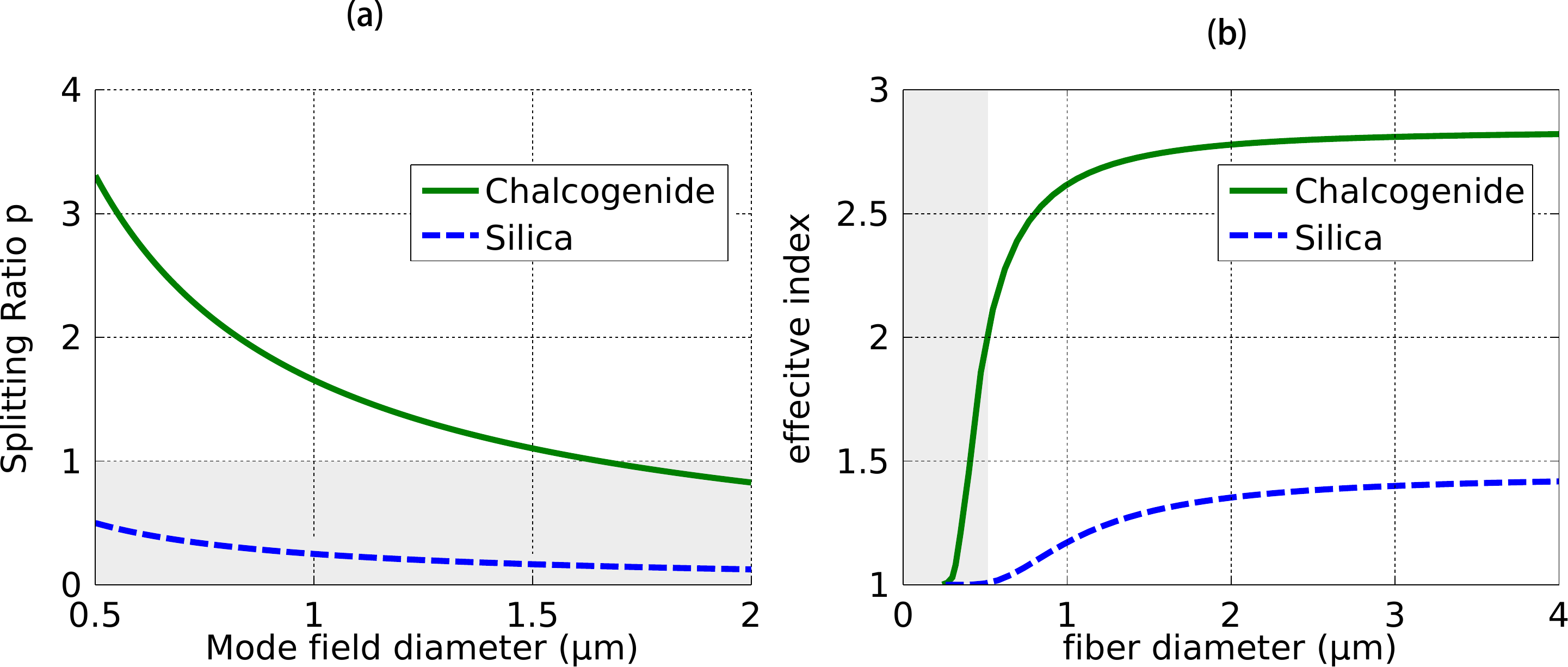}%
 \caption{(a) Splitting ratio $p$ for different mode-field diameters, 50~W pump power and 1 mW Stokes power. \chalco~ and silica fiber taper are compared. Gray area represents the weak coupling region. (b) Effective index calculation for a fiber taper in silica and \chalco~   with respect to the taper diameter. \label{Fig5}}
 \end{figure} 

Once we have seen how the splitting ratio $p$ evolves with the optical power, it is legitimate to investigate the influence of the fiber taper diameter. In our model however, it is more convenient to take into account the mode field diameter (MFD). 
Using this assumption, we compute the splitting ratio for different MFDs and 50~W pump power. The results are summarized in Fig.~\ref{Fig5}(a) for chalcogenide and silica fiber tapers. Once again, the silica fiber taper does not provide enough coupling strength even at submicron scale. This is mainly due to the low effective index of silica fiber taper, as shown in Fig.~\ref{Fig5}(b). For a diameter close to $0.5\mu$m the effective index of the fundamental mode approaches 1, meaning that light is not highly confined into such a low step-index fiber taper. On the other hand, Fig.~\ref{Fig5}(b) shows that the chalcogenide fiber taper has a higher effective index, enabling waveguiding with diameter down to $0.4~\mu$m and a wide range of mode-field diameter to observe the splitting. This is depicted in Fig.~\ref{Fig5}(a) where splitting can be observed with 50~W pump power even with a $1~\mu$m diameter. Moreover, such peak pump power are easily achievable using nanosecond pulses at $1.55~\mu$m and Erbium-doped fiber amplifier (EDFA). As a consequence, we can expect that chalcogenide fiber tapers are best candidate for an experimental demonstration of strong SBS coupling regime.

\section{Conclusion}
In conclusion, we have revisited the theory of stimulated Brillouin scattering in optical waveguides and demonstrated that strong photon-phonon coupling regime can be achieved between the acoustic wave and the two optical waves, giving rise to avoided crossing and Rabi-like splitting. This has been demonstrated by deriving from the coupled equations of SBS a joint system that combines both the phonon and the optical interference pattern that results from the coherent superposition of the pump and Stokes waves. It has been shown that, when the nonlinear coupling rate becomes comparable to the phonon decay rate, the joint system enters in the strong coupling regime and exhibits two splitted eigenvectors $K_\pm(\Omega)$. This regime was further investigated in detail as a function of the Brillouin gain, the pump power, and the mode-field diameter. As a conclusion, our results show that the strong coupling regime could be observed using backward SBS in a chalcogenide-glass 0.7$\mu$m-diameter optical fiber taper, simply by tuning the probe frequency around the Brillouin frequency shift. A splitting of the Brillouin spectrum should be observed at high pump power.

\section{Acknowledgements}

This work was supported by the OASIS project (ANR-14-CE36-0005-01), the support from the R\'egion de Franche-Comt\'e and the LABEX ACTION program (ANR-11-LABX-0001-01).

\appendix
\section{Normalization}\label{A1}
The fields are normalized so that the modulus $|2\TA\TB^*|$ in Eq. (\ref{eqT}) is the optical power. Thus the effective length $\DD$ can be written as
 \begin{eqnarray*}
 	\DD&=&
	\frac{\int\int |E_S|^2\dd x\dd y \sqrt{\int\int |R|^2\dd x\dd y}}{ \int\int \RR(x,y)E_SE_P^*\dd x \dd y}=\sqrt{\pi}\sigma,
\end{eqnarray*}
where $\sigma=\frac{\mathrm{MFD}}{2}$ is the waist and MFD is the mode field diameter (after Ref.\cite{MFD}). We can also define the effective mode area as $A_{eff}=D_{eff}^2$.

\section{Group velocity of the optical interference pattern}\label{A2}

If we do not neglect the group velocity in Eq. (\ref{eqT}), we get
    \begin{eqnarray}  
	\frac{\partial \TA_{P}\TB_{S}^* }{\partial z}
	 +\frac{\partial \beta}{\partial \omega}
	 \left\{
	\frac{\partial \TA_{P}}{\partial t}\TB_{S}^*
	-\TA_{P}\frac{\partial\TB_{S}^*}{\partial t}	 
	 \right\}
	 &=& 
	  -\frac{\jj \omega_P\omega_S\gamma_e\rr}{2n \cc \rho_0\DD} 
	 \left(\frac{\TA_P\TB_{P}^*}{\omega_P}-\frac{\TA_S\TB_{S}^*}{\omega_S}\right).\nonumber
  \end{eqnarray} 
Note that if the two P and S waves were propagating along the same direction, the left-hand side (LHS) would become $\frac{\partial \TA_{P}\TB_{S}^* }{\partial z}+\frac{\partial \beta}{\partial \omega}\frac{\partial \TA_{P}\TB_{S}^*}{\partial t}$, giving rise to a forward optical beating travelling at the mean group velocity. In the counterpropagating case, it is much more difficult to define a group velocity for the optical interference pattern. 
However, we can show that the \textit{effective} group velocity is somehow bounded and has a negligible effect on our main results. 
We can see in Fig.~\ref{Fig3}(a) three different scenario. First, if the pump is a continuous wave, $\frac{\partial \TA_{P}}{\partial t}=0$, and the Stokes wave is an optical pulse, the intensity fluctuations due to the superposition of both waves is located at the same location as the Stokes pulse. The optical beat pattern thus follows the Stokes pulse and propagates at the same group velocity. It translates into the equation by the fact that $\frac{\partial \TA_{P}}{\partial t}=0$. Note that the interaction with the phonon is not likely to disturb this equilibrium if the pump is sufficiently strong compared to the Stokes wave $|\TA_P|\gg |\TA_S|$. Fig.~\ref{Fig3}(b) shows the opposite scenario where the Stokes is a continuous wave and the pump a pulse. In this case, the optical beating follows the pump pulse and its group velocity is therefore the pump group velocity. The striking consequence is that this group velocity is the opposite of the previous case, as Pump and Stokes propagate in opposite direction. From the equation, we can address a third interesting example if $\TA_P=\TB_P^*$, the group velocity of the optical beating is then null. This is depicted in Fig~\ref{Fig3}(c), where both pump and Stokes waves are optical pulses and that the barycenter of both pulse is static. This last case is however very unlikely to happen since an increasing of $\TB_S^*$ over time due to the phonon is clearly related to a depletion of the Pump $\TA_P$ with much impact on the group velocity. 
    \begin{figure}[hbt]
  \centering
 \includegraphics[width=\columnwidth]{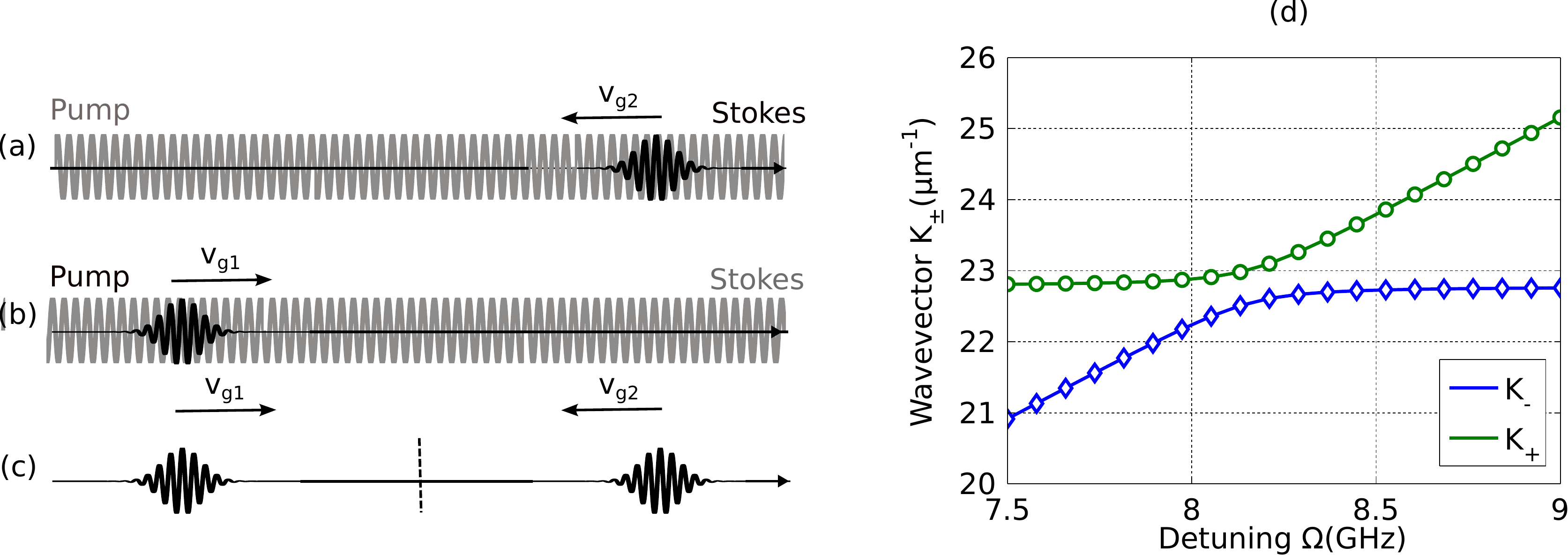}%
 \caption{(a) Pump in gray is a continuous wave and travels forward the z-axis, Stokes pulse light travels backward. (b) Pump pulse in black travels forward the z-axis, Stokes light in gray is continuous and travels backward. (c) Both lights are pulses and travel in opposite direction. The resulting wave packet is static. (d) Same computation as in Fig.~\ref{Fig3} but with $\frac{1}{v_g}=-\frac{\partial \beta}{\partial \omega}$ in plain green and blue and with $\frac{1}{v_g}=+\frac{\partial \beta}{\partial \omega}$ in markers \label{Fig6}}
 \end{figure} 
 As a result, if the group velocity of the optical interference pattern may be uncertained, it is certainly bounded by the relation $\frac{1}{v_g}\in \left[ -\frac{\partial \beta}{\partial \omega},+\frac{\partial \beta}{\partial \omega}\right]$ where the limits correspond to the two first cases. More intriguingly, thought these boundaries are wide appart, they do not affect our main results. Figure~\ref{Fig6}(d) shows a similar numerical simulation of the avoided crossing as in Fig.~\ref{Fig3}, but with  $\frac{1}{v_g}=-\frac{\partial \beta}{\partial \omega}$ in plain green and blue and $\frac{1}{v_g}=+\frac{\partial \beta}{\partial \omega}$ with markers. There is no noticeable difference because $\left |  \frac{1}{v_g}\right |\simeq \left |  \frac{n}{\cc}\right |\ll  \left |  \frac{1}{v_a}\right |$ is very small compared to $ \left |  \frac{1}{v_a}\right |$ and thus negligible whatever the sign is.
The second point we want to address is that in most Brillouin experiments, the group velocity is always positive since the pump power is greater than the Stokes one before depletion. Such assumption leads to $	\left|\frac{\partial \TA_{P}}{\partial t}\TB_{S}^*\right| \ll \left|\TA_{P}\frac{\partial\TB_{S}^*}{\partial t}\right|$ leading to a positive group velocity.

\section{Brillouin gain}\label{A3}
The Brillouin gain was defined using the method described in \cite{Agrawal}. From the above equations (\ref{eq:Y1}) and (\ref{eq:Y4}), we find
  \begin{eqnarray*}  
	\frac{\partial \TA_{P}\TB_{P}^* }{\partial z}
	 +\frac{\partial \beta}{\partial \omega}\frac{\partial \TA_{P}\TB_{P}^*}{\partial t}
	 &=& \frac{\jj \omega_P\gamma_e}{2n_P \cc \rho_0} \frac{1}{\DD}\left(\rr\TA_S\TB_{P}^*
	+\rr^*\TA_P\TB_S^*\right) ,\\
	-\frac{\partial \TA_{S}\TB_{S}^* }{\partial z}
	 +\frac{\partial \beta}{\partial \omega}\frac{\partial \TA_{S}\TB_{S}^*}{\partial t}
	 &=&\frac{\jj \omega_S\gamma_e}{2 n_S\cc \rho_0}\frac{1}{\DD}
	 \left(\rr^*\TA_P\TB_S^*+\rr\TA_S\TB_P^*\right).\\
%
  \end{eqnarray*} 
These equations first show that the optical powers propagate at the same group velocity. Moreover, if we neglect the nonlinear coupling term and then substract both equations, we get the Poynting theorem for plane waves, $\ddiv \VS=0$ and the conservation law $\frac{\partial \TA_{S}\TB_{S}^*}{\partial t}-\frac{\partial \TA_{P}\TB_{P}^*}{\partial t}=0$, or $\frac{\partial P_S}{\partial t}+\frac{\partial P_P}{\partial t}=0$.

Using the phonon equation (\ref{eq:ph}) in steady-state regime, in order to substitute $\rr$, we find

     
       \begin{eqnarray*} 
	\rr(z,t) 
	&=&
	-\frac{1}{\Omega_B^2-\Omega^2-\jj\Omega\Gamma_B}
%
	\frac{\gamma_e\beta_a^2}{n\cc}\frac{1}{\DD}
	\TA_{P}\TB^*_{S}, 
     \end{eqnarray*}   
     and assuming $\omega_P\simeq\omega_S$ we thus obtain the usual power equation evolution of SBS, as in Ref.\cite{Agrawal}
       \begin{eqnarray*} 
	\frac{\partial P_P }{\partial z}
	&=&
	-\frac{g}{A_{eff}}P_PP_S\\
	\frac{\partial P_S }{\partial z}
	&=&
	\frac{g}{A_{eff}}P_PP_S\\
     \end{eqnarray*}        
 where $A_{eff}=\DD^2$, $\kappa_1=\frac{ \omega_S\gamma_e}{2 n_S\cc\rho_0}$ and $\kappa_2=	\frac{\gamma_e\beta_a^2}{2n\cc}\frac{1}{\Omega}=\frac{\gamma_e \omega_S}{\cc^2 v_a}$    
    with  
 
\begin{eqnarray*} 
%
	g(\Omega)&=&
	g_0
	\frac{\left(\Gamma_B/2\right)^2}{(\Omega_B-\Omega)^2+\left(\Gamma_B/2\right)^2},
\end{eqnarray*}        
where $g_0=4\kappa_1\kappa_2\frac{1}{\Gamma_B}=2\frac{ \omega_S\omega_P\gamma_e^2}{ n_S\cc^3v_a\rho_0\Gamma_B}$ \cite{Agrawal}.

\section{Glass parameters}\label{A4}
In our simulations, we used and compared silica-based to chalcogenide \chalco ~glass-based fiber tapers, because they are readily available and manufactured \cite{Baker2012}. Moreover, recent experiments have shown great potential of these glass materials for SBS applications \cite{BeugnotOL2014,Beugnot2014}. The table below summarizes the optical and acoustical parameters of these two materials. Note that the \chalco~Brillouin linewidth is taken larger than the one in \cite{Chalco} because of the spectral broadening due to fiber coating \cite{BeugnotOL2014}.
 \begin{table}[h]
\caption{\label{tab1} Table of optical and acoustical parameters of silica-based and chalcogenide-based glass materials.}
\begin{indented}\item[]
\footnotesize
\begin{tabular}{@{}lll}
\br
 	 & silica$^{a}$ & \chalco$^{a}$\\
\mr
 refractive index at 1.5~$\mu$m & 1.45 & 2.8\\
 Brillouin frequency shift (GHz)  & 11 & 8\\
  Brillouin linewidth (MHz)  & 30 & 50\\ 
 Acoustic velocity (ms$^{-1}$) & 5600 & 2250\\ 
 Brillouin gain ($\mathrm{m}\mathrm{W}^{-1}$)& $3.3.10^{-11}$ & $600.10^{-11}$\\
Mean mass density (Kg.m$^{-3}$) & 2200 & 4640\\ 
\br
\end{tabular}\\
$^{a}$See \cite{Agrawal}; $^{b}$See \cite{Chalco}.
\end{indented}
\end{table}
\normalsize

\newpage

\section{Anti-Stokes scattering}\label{A5}

\begin{figure}[hbt]
  \centering
 \includegraphics[width=0.33\columnwidth]{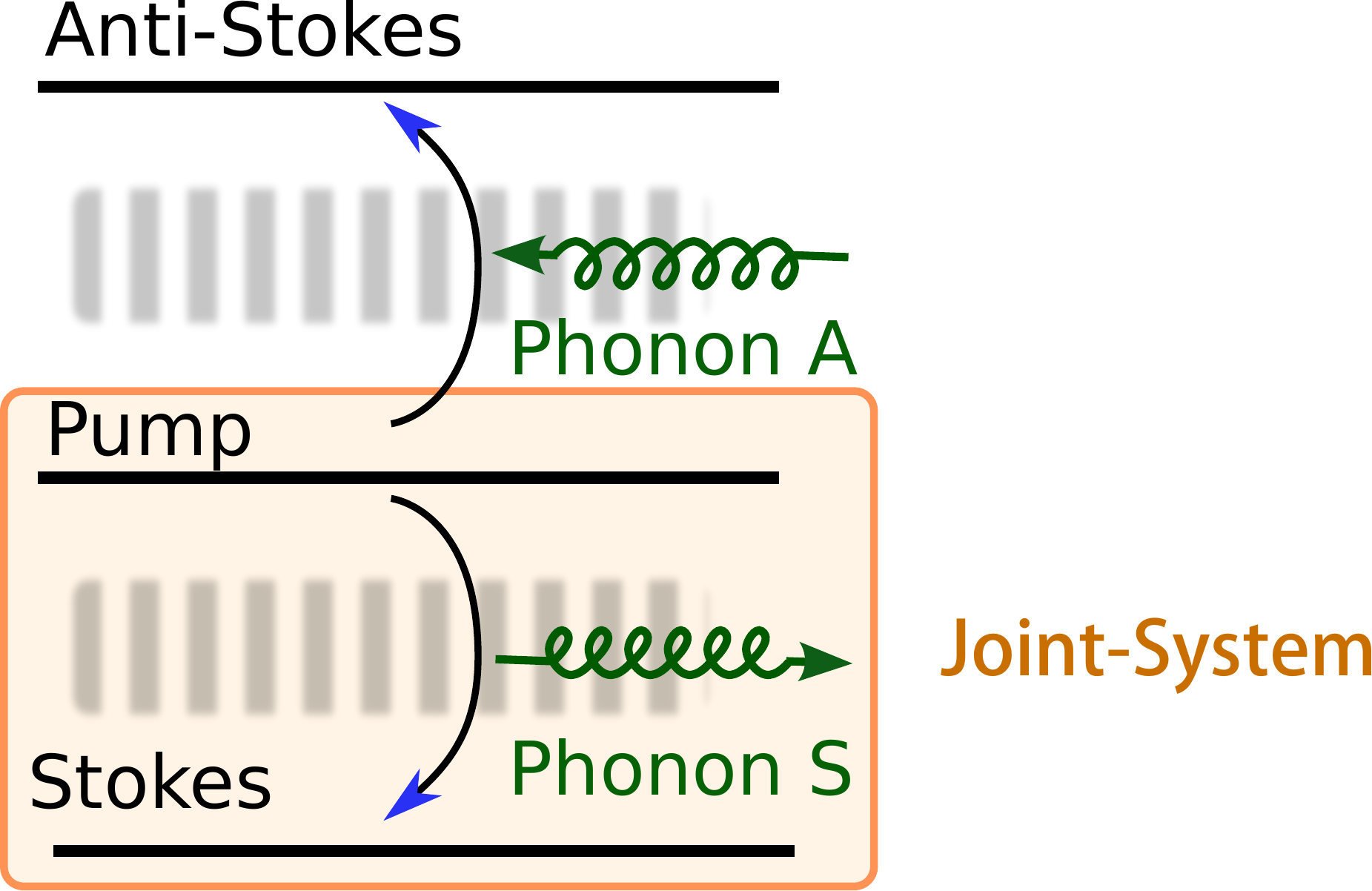}%
 \caption{Scheme of the anti-Stokes scattering process above the joint pump-Stokes system.\label{Fig7}}
 \end{figure}

One of the main hypothesis for the strong coupling Brillouin regime was to consider both the anti-Stokes and second-order Stokes scattering as negligible. This assumption is valid in the weak pump to Stokes conversion regime. However, adding the anti-Stokes wave as a higher energy level in the system equation does not significantly change our results. As shown in Fig.~\ref{Fig7}, anti-Stokes scattering involves an oncoming acoustic phonon $A$, that must not be mistaken for the forward propagating phonon $B$, involved in the joint pump-Stokes system\cite{Boyd}. The scheme illustrates that anti-Stokes scattering can then be considered as an additional loss for the pump level. As a result, the model could be considered as a 3-level system\cite{Haroche}. The pump depletion due to this phenomenon would then be associated to a decay rate related to the room temperature phonon population. However, in optical fibers, anti-Stokes scattering is a much weaker than Stokes scattering due to the low phonon population and the fast phonon decay rate. Therefore, it is usually neglected and particularly in the stimulated regime \cite{Kobyakov}. In the strong coupling regime, the coupling must be stronger than the phonon decay that is around few microns. Over such a short distance, the pump decay rate associated with anti-Stokes scattering is thus negligible. Note that second-order Stokes scattering can be neglected for the same reasons.

\end{document}